\documentclass[12pt]{article}
\usepackage[super,compress]{cite}
\usepackage{amsmath}
\usepackage{amssymb}
\usepackage{graphicx}
\usepackage{float}
\begin{document}

\begin{center}
{\bf
High energy nucleus-nucleus collisions with accounting for Coulomb
interaction.}

Yu.M. Shabelski and A.G. Shuvaev \\

\vspace{.5cm}

Petersburg Nuclear Physics Institute, Kurchatov National
Research Center\\
Gatchina, St. Petersburg 188300, Russia\\
\vskip 0.9 truecm
E-mail: shabelsk@thd.pnpi.spb.ru\\
E-mail: shuvaev@thd.pnpi.spb.ru

\end{center}

\vspace{1.2cm}

\begin{abstract}
\noindent
The differential elastic cross sections of
$^{12}$C - $^{12}$C, $^{16}$O - $^{16}$O and
$^{20}$Ne - $^{20}$Ne
nuclei scattering are calculated in the complete Glauber theory
with the account of the modification due to Coulomb interaction
and form factor effects.
The role of the Coulomb interaction is shown to be significant
mainly in the diffractive minima.
The results of the complete Coulomb calculations are
noticeably different from those obtained in the Born approximation.
\end{abstract}

\section{Introduction}
Investigation of high energy nucleus-nucleus scattering
provides an important information on the colliding nuclei
structure. In particular, the exotic nuclei have been found,
which contain one or more nucleons in the so-called halo
configuration~\cite{Ozawa:2000gx}.
The correct analysis of the scattering of the charged
object such as nuclei needs the consistent treatment
of the interplay of the strong and Coulomb interactions.
The Coulomb effects determined by the product
of the electromagnetic coupling and the electric
charges of the colliding nuclei, $\alpha_{em}Z_A Z_B$,
can be significant.
An essential difficulties come from the infinite range of
the Coulomb forces that causes the infrared divergency
and calls for an appropriate cutoff.
Besides, as the electric charges are not
point like but spread over the nuclear volume, a more detailed analysis
is required.

One has to point out that the Coulomb and the strong interaction
interference has been mainly studied for the proton-proton
or proton-nuclear scattering rather than for the nucleus-nucleus
scattering. The latter case is hampered by far more complex structure
the Glauber theory of the strong interactions has
beyond the optical approximation.

In this paper we consider the elastic scattering of relatively
light nuclei, $^{12}$C--$^{12}$C, $^{16}$O--$^{16}$O,
$^{20}$Ne--$^{20}$Ne.
For the comparison we also present the results for the elastic
proton scattering on $^{12}$C.

To avoid the infrared divergencies showing up at the small
scattering angles we deal with the differential cross section
for not very low transfer momentum $Q^2$.
The Coulomb effects turn out to be most sizable at
the diffraction minima and increasing with the atomic number.
The difference between the complete Coulomb calculation made
in the eikonal approach and the Born approximation appears
to be quite noticeable.

\section{Nucleus-nucleus elastic scattering}
The elastic scattering amplitude of two nuclei $A$ and $B$
at comparatively
high energy (about 1~GeV per nucleon) is given by the Glauber formula
\begin{equation}
\label{bint}
f_{AB}(q)\,=\,\frac{ip}{2\pi}\int d^{2}b\,e^{i qb}\,
\bigl[1\,-\,s_{AB}(b)\bigr],
\end{equation}
where $p$ is a relative momentum in the central of mass frame,
$q$ is the transferred momentum. The impact parameter $b$ is
a two dimensional vector
in the transverse plane with respect to relative momentum
of the colliding nuclei $A$, $B$.
The evaluation of the function $s_{AB}(b)$
relies on the short range of the strong
interaction.
Due to this property the phase shift on a nucleus
comes out the sum of those for the independent
scattering of the constituent nucleons.
Denoting by $f_{NN}(q)$ the nucleon-nucleon scattering amplitude
and by $s_{NN}(b)$ the phase shift,
$$
\Gamma_{NN}(b)\,\equiv\,
1\,-\,s_{NN}(b)\,=\,\frac 1{2\pi i p}\int d^2q\,e^{iqb}f_{NN}(q),
$$
we have
\begin{equation}
\label{I}
s_{AB}(b)\,=\,\langle\,A,\,B\,|
\left\{\prod\limits_{i\,j}\bigl[1-\Gamma_{NN}(b+x_i-y_j)\bigr]
\right\}
|\,A,\,B\,\rangle,
\end{equation}
where the brackets stand for an average over the nucleons' positions
$x_i$ and $y_j$ lying in the same plain with the impact
parameter.

There is much study addressing the interplay of the strong
and Coulomb interactions in the context
of eikonal approach
\cite{Bethe, Islam:1967zz, West:1968du, Cahn, Kopeliovich}
(and references therein).
In this paper we adopt to the nucleus-nucleus collision
the method proposed by
Glauber~\cite{Glauber:1970jm, Glauber:2019roq}
for the proton-nucleus case.
The strong amplitude is modified by adding the Coulomb phase
\begin{equation}
\label{sC}
s_{AB}(b)\,\to\,s_{AB}^C(b)\,=\,s_{AB}(b)e^{i\chi_C(b)},
\end{equation}
which is evaluated as the eikonal factor
\begin{equation}
\label{ChiC}
\chi_C(b)\,=\,-\frac{M_{AB}}{p}\int_{-\infty}^\infty
dz\,V_c(z,b),
\end{equation}
where $M_{AB}$ is reduced mass and
$$
V_c(r)\,=\,Z_A Z_B\alpha_{em}
\int d^3r_1 d^3r_2\,
\frac{\rho_A(r_1)\rho_B(r_2)}{|r+r_1-r_2|}.
$$
is the Coulomb potential
experienced by the colliding nuclei A and B,
$\rho_{A,B}(r)$ being their densities distributions,
$\int d^3r\,\rho_{A,B}(r)\,=\,1$.

The long range character of the Coulomb interaction
makes the integral (\ref{ChiC}) divergent
and requires the infrared cutoff.
Taking the fictitious photon mass $m_{ph}$ as a regulator
the infinite part is separated out as
$$
\chi_C(b)\,=\,\gamma\ln\frac 14  b^2m_{ph}^2\,+\,\chi_C^\prime(b),
~~~\gamma\,\equiv\,Z_A Z_B\alpha_{em}\frac{M_{AB}}{p}
$$
with a remainder $\chi_C^\prime(b)$ finite for $m_{ph}\to 0$.

To avoid the infrared cutoff dependence
we proceed in a way proposed
in~\cite{Glauber:2019roq} (see also~\cite{Petrov:2020tnr,
Durand:2020fbt} for further implementations).
Rewriting the amplitude~(\ref{bint}) as
\begin{eqnarray}
f_{AB}(q)\,&=&\,2\pi i p\delta^{(2)}(q)
-e^{i\gamma\ln \frac 12 m_{ph}^2\lambda^2}\frac{ip}{2\pi}
\int d^{2}b\,e^{i qb}\,e^{i\ln\frac{b^2}{2\lambda^2}}\,
e^{i\chi_C^\prime(b)}s_{AB}^C(b) \nonumber \\
&\equiv&\,2\pi i p\,\delta^{(2)}(q)\,+\,
e^{i\gamma\ln \frac 12 m_{ph}^2\lambda^2}
\frac{ip}{2\pi}\varphi_{AB}^\prime(q),\nonumber
\end{eqnarray}
where $\lambda$ is a typical size of the nucleus
density distribution (see eq.(\ref{HO}) below),
the first term is irrelevant for non vanishing
momentum transfer. It drops out of the amplitude
by staying away from extremely small scattering angles.
In this domain the amplitude amounts to the second term
where the overall $q$-independent
phase factor in front can be stripped off,
\begin{equation}
\label{fp}
f_{AB}^\prime(q)\,=\,\frac{ip}{2\pi}\varphi_{AB}^\prime(q).
\end{equation}
It is convenient to rearrange it further to the form
\begin{eqnarray}
\label{phi}
\varphi_{AB}^\prime(q)\,&=&\,
-\int d^{2}b\,e^{i qb}\,e^{i\gamma\ln\frac{b^2}{2\lambda^2}}
e^{i\chi_C^\prime(b)}s_{AB}(b)  \\
&=&\,\int d^{2}b\,e^{i qb}\,e^{i\gamma\ln\frac{b^2}{2\lambda^2}}
e^{i\chi_C^\prime(b)}\bigl[1-s_{AB}(b)\bigr]
+\int d^{2}b\,e^{i qb}\,e^{i\gamma\ln\frac{b^2}{2\lambda^2}}
\bigl[1-e^{i\chi_C^\prime(b)}\bigr] \nonumber \\
&&\,-\int d^{2}b\,e^{i qb}\,e^{i\gamma\ln\frac{b^2}{2\lambda^2}}.
\end{eqnarray}
The last integral here is analytically evaluated to
$$
\int d^{2}b\,e^{i qb}\,
\biggl(\frac{b^2}{2\lambda^2}\biggr)^{i\gamma}\,=\,
4\pi\frac{\Gamma(1+i\gamma)}{\Gamma(-i\gamma)}\,
\frac 1{q^2}\bigl(\frac 12 q^2\lambda^2\bigr)^{i\gamma}.
$$

The complete Glauber strong
amplitude~(\ref{I}) has been evaluated
using the generating function method proposed in our previous
paper~\cite{Shabelski:2021iqk},
\begin{equation}
\label{ddZ}
s_{AB}(b)\,=\,\frac 1{Z(0,0)}
\frac{\partial^A}{\partial u^A}
\frac{\partial^B}{\partial v^B}\,
Z(u,v)\biggl|_{u=v=0},
\end{equation}
with the generating function expressed through the nucleon
distributions $\rho_{A,B}(x)$ in the colliding nuclei,
\begin{eqnarray}
\label{Zuv}
Z(u,v)\,&=&\,e^{W_y(u,v)},~~~~~~~~
z_y\,=\,1-\frac 12 \frac{\sigma_{NN}^{tot}}{a^2},~~~
\rho_{A,B}^\bot(x_\perp)\,=\,
\int d z\,\rho_{A,B}(z,x_\perp),
\\
\label{Wy}
W_y(u,v)\,&=&\, \frac 1{a^2}\int d^{\,2}x\,
\ln\bigl(\!\!
\sum\limits_{M\le A,N\le B}
\frac{z_y^{M\, N}}{M!N!}
\bigl[a^2 u\rho_A^\bot(x-b)\bigr]^M
\bigl[a^2 v\rho_B^\bot(x)\bigr]^N
\bigr).
\end{eqnarray}
Here $\sigma_{NN}^{tot}$ is the total nucleon-nucleon
cross section, the effective interaction radius $a^2 = 2\pi\beta$
is related to the slope of the nucleon-nucleon elastic
amplitude
$$
f_{NN}^{el}(q)\,=\,ik\frac{\sigma_{NN}^{tot}}{4\pi}
e^{-\frac 12\beta q^2},
$$
$k$ is the mean momentum carried by a nucleon in the incident nucleus.
The nucleon density is parameterized by
harmonic oscillator distribution well suited for $A\le 20$,
\begin{equation}
\label{HO}
\rho_A(r)\,=\,\rho_0\bigl[1+\frac 16\,(A-4)\frac{r^2}{\lambda^2}
\bigr]e^{-\frac{r^2}{\lambda^2}},
\end{equation}
the parameter $\lambda$ being adjusted to the mean square radius.
With this density the generating function is closely expressed
in terms of the overlap functions
$t_{mn}(b)=\int d^2x[\rho_A^\bot(x-b)]^m [\rho_B^\bot(x)]^n$
(see \cite{Shabelski:2021iqk} for details).
The obtained Glauber amplitude~(\ref{ddZ}) has to be plugged
in the final amplitude~(\ref{phi}) along with
the eikonal factor $\chi_C^\prime(b)$ calculated for
the density distribution~(\ref{HO}).
The differential cross section reads
\begin{equation}
\label{dsig}
\frac{d\sigma}{dt}\,=\,\frac{\pi}{p^2}
\bigl|f_{AB}^\prime(q)\bigr|^2,
~~~t=-q^2\not = 0.
\end{equation}
\begin{figure}[H]
\hskip 1 cm
\includegraphics[width=.4\hsize]{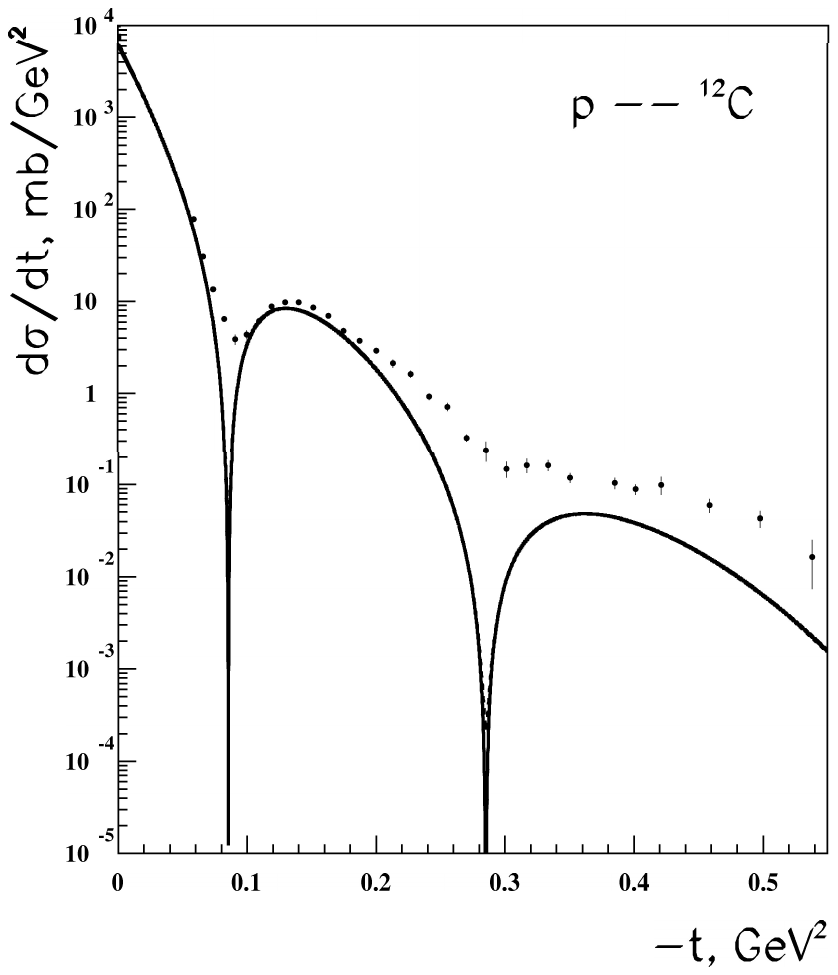}
\includegraphics[width=.4\hsize]{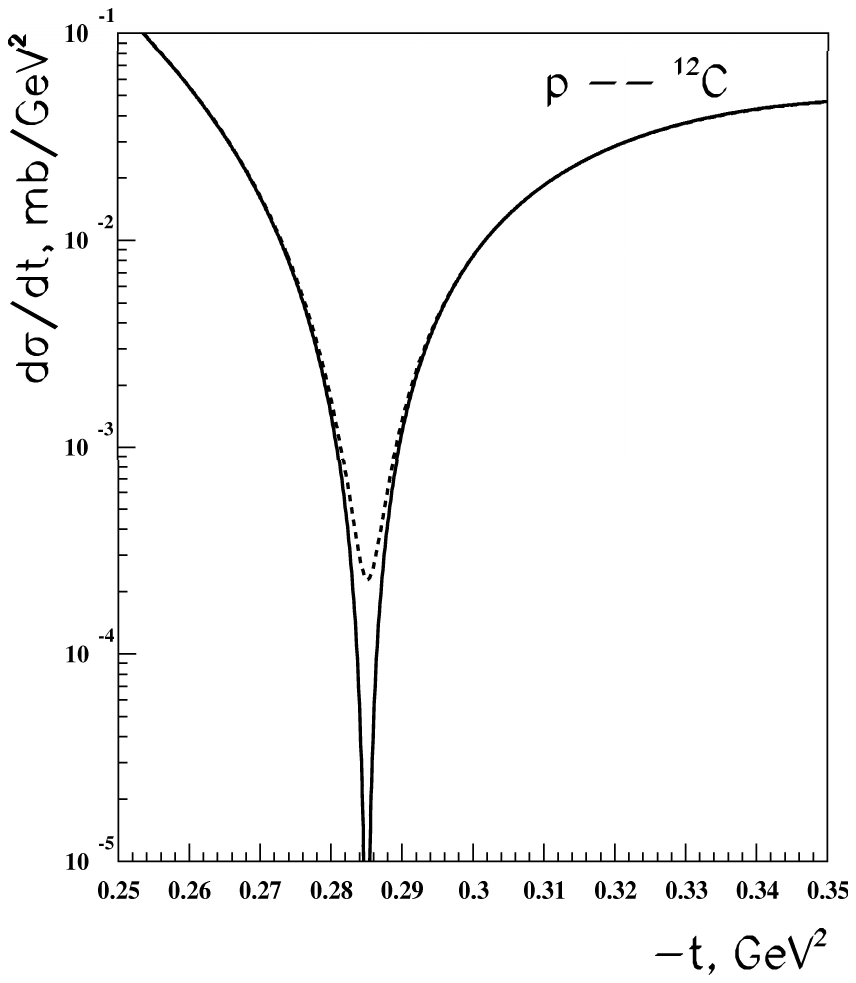}
\vskip -1.cm \caption{\footnotesize
Left: Differential cross section of 1~GeV proton scattering
on $^{12}$C target calculated in the Glauber theory
without, solid line, and
with account for the Coulomb phase (\ref{phi}), dashed line.\\
Right:
Enlarged fragment of the left panel near the second diffractive
minimum.\\
The experimental points are taken from~\cite{Alkhazov:1972oie}.\\
}
\end{figure}

\section{Results of calculations}
Here we consider the collision of the projectile nucleus with atomic number $B$ and
mass $M_B=B m_N$ on the target nucleus with atomic number $A$ and mass $M_A=A m_N$,
$m_N$ is the nucleon mass. The scattering amplitude is determined in the central of
mass frame by the reduced mass $M_{AB}= M_A M_B/(M_A+M_B)$ and the relative
momentum $p$. The projectile energy is characterized by the kinetic energy per
nucleon in the laboratory frame. The presented below results are obtained for the
scattering energy about 1~CeV per nucleon.

\begin{figure}[H]
\includegraphics[width=.4\hsize]{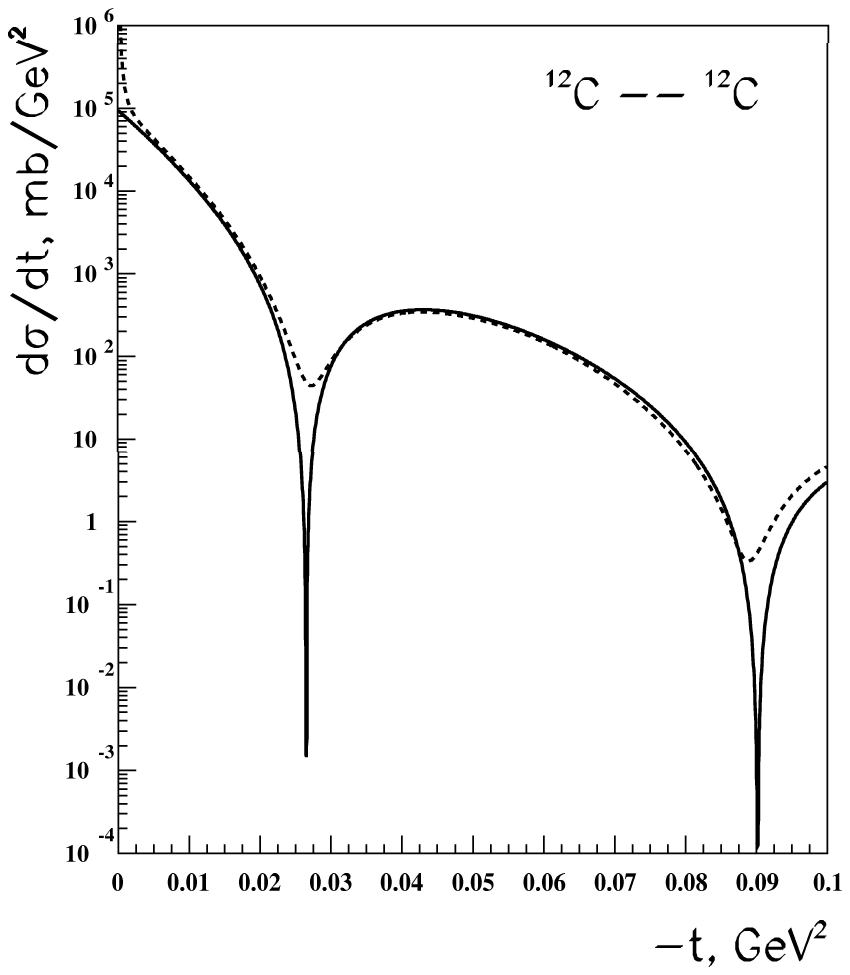}
\hskip -2cm
\includegraphics[width=.4\hsize]{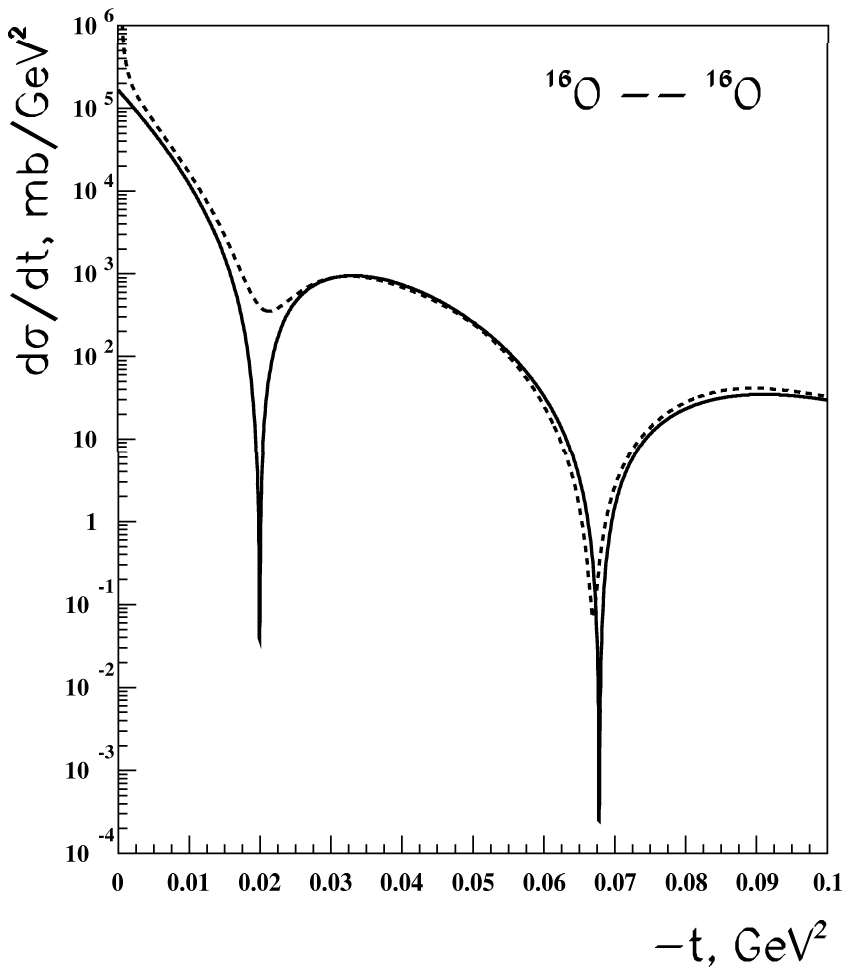}
\hskip -2cm
\includegraphics[width=.4\hsize]{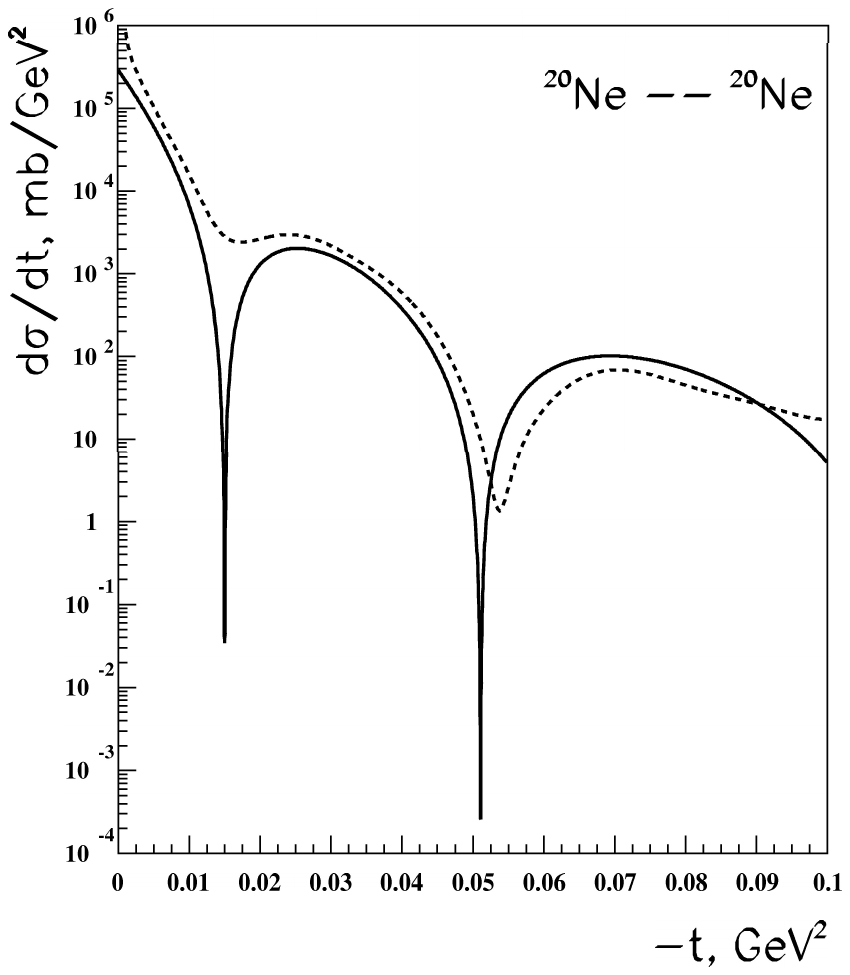}
\vskip -1.cm
\caption{\footnotesize
Differential cross section of $^{12}$C - $^{12}$C, left,
$^{16}$O - $^{16}$O, central, and $^{20}$Ne - $^{20}$Ne, right,
scattering at the energy 1~GeV per nucleon
calculated with the pure Glauber expression (\ref{Zuv}), solid lines,
and with Coulomb phase (\ref{phi}), dashed line.
}
\end{figure}
We start out by reproducing the experimental data
on elastic scattering of 1~GeV protons
on $^{12}$C target~\cite{Alkhazov:1972oie}.
The calculation has been done with the distribution~(\ref{HO})
for $^{12}$C nucleus and point like density for the proton.
The generating function amounts in this case
to the so called rigid target
approximation (see \cite{Shabelski:2021iqk}).
The results of the calculations are presented in Fig.~1
together with the experimental points taken
from~\cite{Alkhazov:1972oie}. They show the Coulomb contribution
to be somewhat distinguishable subject to the pure Glauber cross
section only in the diffractive minima.
The discrepancy with the experiment can be explained probably
by the oversimplified form of the nuclear density~(\ref{HO})
as well as by
non accountable parts of nucleon-nucleon amplitude.
(see discussion in~\cite{Alkhazov:1978et}).

\begin{figure}[H]
\includegraphics[width=.4\hsize]{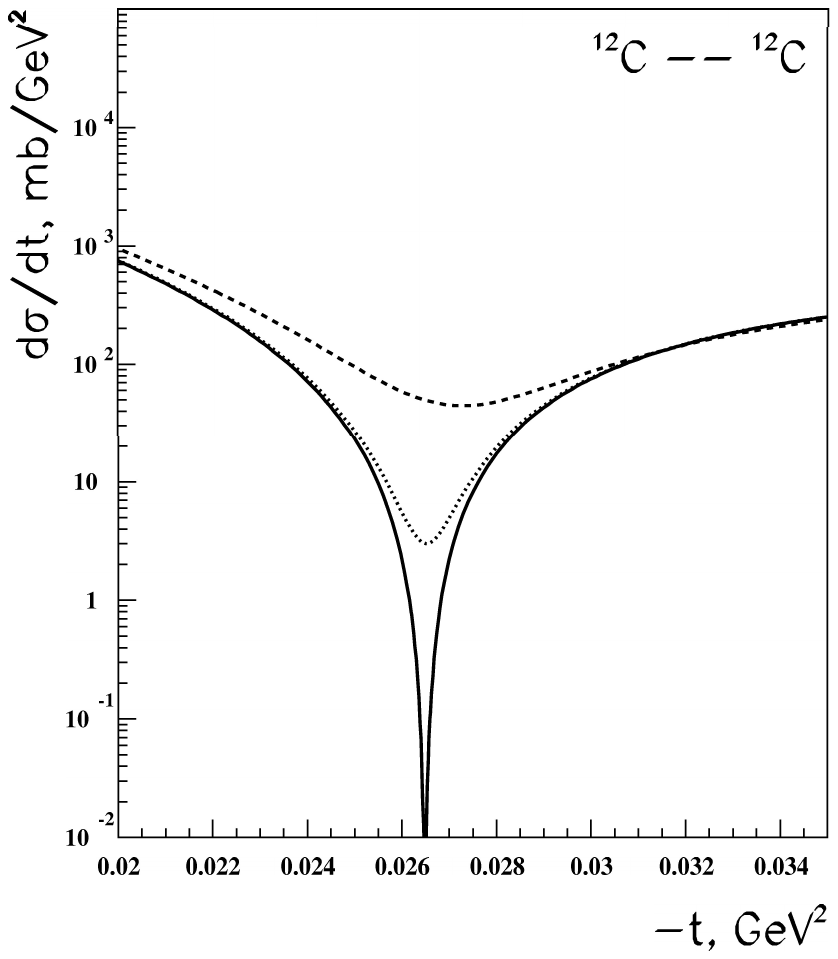}
\hskip -2cm
\includegraphics[width=.4\hsize]{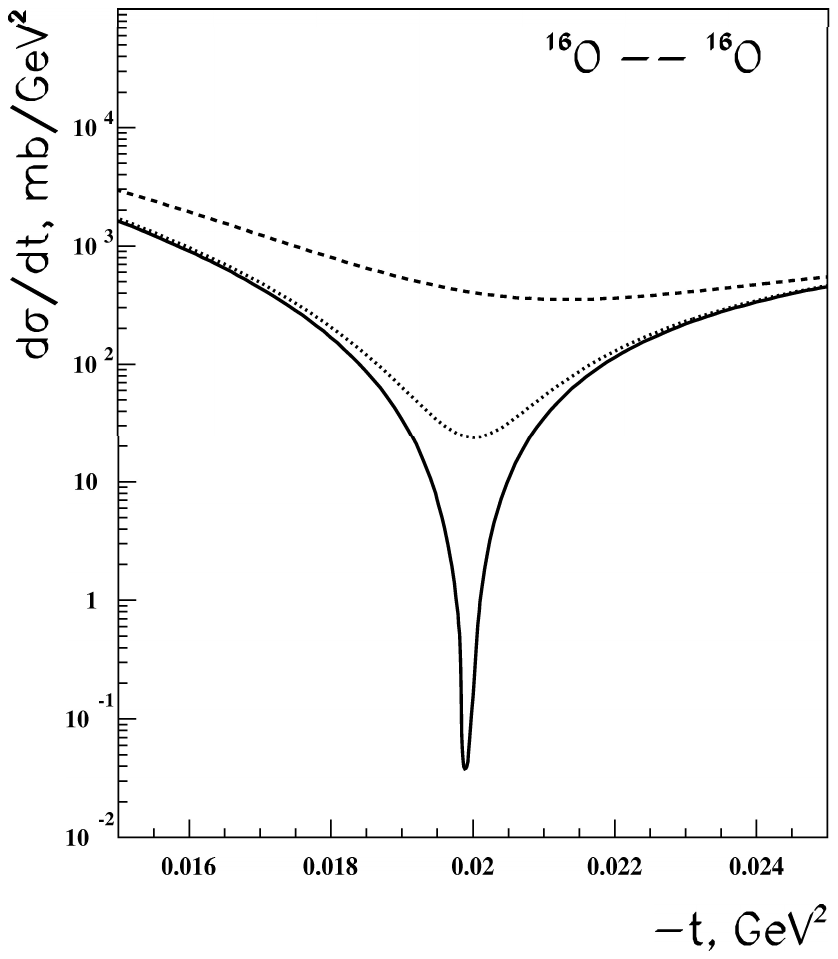}
\hskip -2cm
\includegraphics[width=.4\hsize]{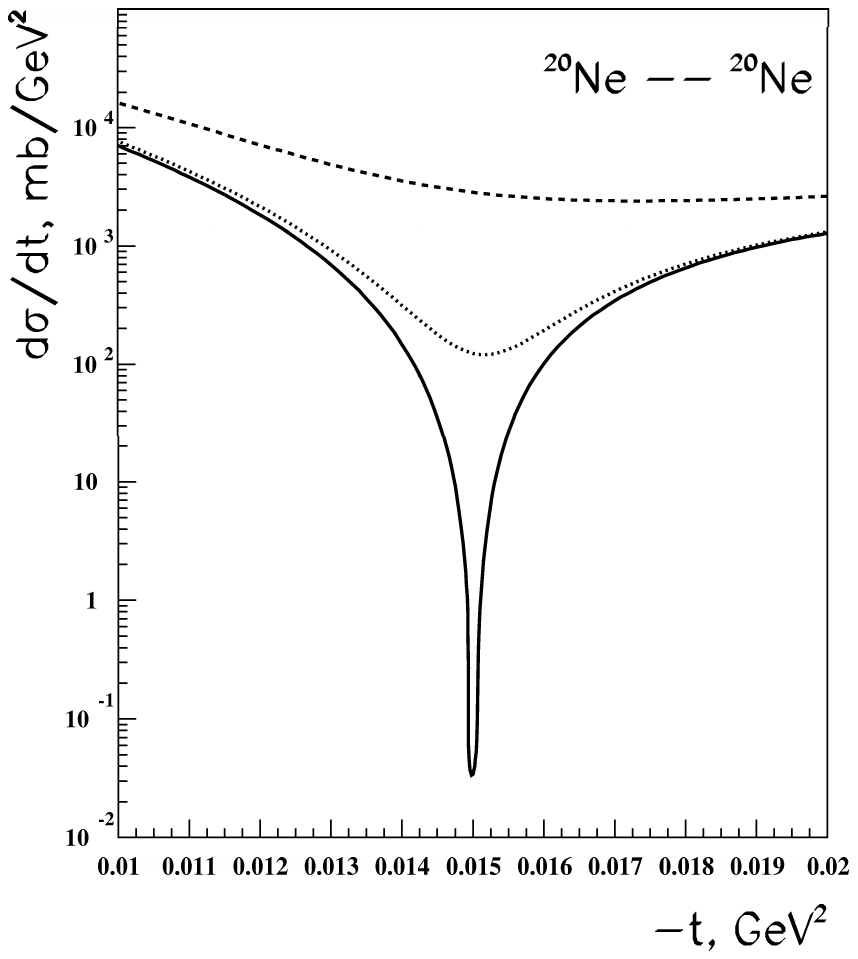}
\vskip -1.cm
\caption{\footnotesize
Differential cross section of $^{12}$C - $^{12}$C, left,
$^{16}$O - $^{16}$O, central and
$^{20}$Ne - $^{20}$Ne, right, scattering
at the energy 1~GeV per nucleon
calculated without the Coulomb corrections, solid line,
with the first order Born Coulomb contribution
added to the Glauber amplitude, Eq.~(\ref{Born}), dotted line
and with the phase factor Eq.~(\ref{phi}), dashed line.
}
\end{figure}

Next we go to the nucleus-nucleus scattering taking three pairs
of the heavy ions, $^{12}$C - $^{12}$C, $^{16}$O - $^{16}$O and
$^{20}$Ne - $^{20}$Ne with $Z_A Z_B = $ 36, 64 and 100.
The parameter $\lambda$ in the densities~(\ref{HO}) has been chosen
to fit the mean square radii $R_{ms}=$2.49~fm, 2.69~fm
and 2.99~fm \cite{Alkhazovi:2011ty}.
The results for the energy 1~GeV per nucleon are in Fig.~2.
The Coulomb contribution is found to be relatively larger
in all three cases than in p - $^{12}$C scattering.
When the product $Z_A Z_B$ gets larger the Coulomb contribution
increases,
which is well demonstrated by $^{20}$Ne - $^{20}$Ne scattering
where the essential difference with the pure Glauber cross section
is seen after the first diffractive minimum.

These results are compared to those given by the first
order Coulomb contribution,
in which the interaction
is approximated with the Born term added
to the strong amplitude,
\begin{equation}
\label{Born}
\frac{d\sigma}{dt}\,=\,\frac{\pi}{p^2}
\bigl|f_C(q)\,+\,f_{AB}(q)\bigr|^2,
~~~f_C(q)\,=\,-2M_{AB}Z_A Z_B e^2\,\frac{\rho_A(q)\rho_B(q)}{q^2},
\end{equation}
$$
\rho_{A,B}(q)\,=\,\int d^3x\,e^{iqx}\rho_{A,B}(x).
$$
The cross section (\ref{Born}) turns out to be different
from the pure Glauber one
only near the first diffractive minimum.
Fig.3 shows that
even there it is much smaller than that resulting from
the full Coulomb phase.

\section{Conclusion}
We have calculated the differential elastic cross sections of
the nucleus-nucleus scattering. The Coulomb interaction has been
included through the extra phase factor (\ref{sC}) inserted into
the Glauber amplitude. It obviously does not change the reaction cross section,
that is the difference between the total cross section and the integrated
elastic one,
$$
\sigma_{AB}^{r}\,=\,\sigma_{AB}^{tot}\,-\,\sigma_{AB}^{el}
\,=\,
\int\!\! d^2b\,\bigl[1\,-\,|s_{AB}(b)|^2\bigr],
$$
though it affects these two cross sections taken individually as well as the
differential cross section. The long range nature of the Coulomb interaction causes
the infrared divergency. However as the divergent pieces are peaked about the small
scattering angles the differential cross section comes out to be finite for not too
small momentum transfer.

It is worth to point out that the adopted here treatment
of the Coulomb contribution as a pure phase implies only
elastic effects experiencing by nuclei. There are no
way to describe in this manner the Coulomb excitations
like the giant resonance, nuclear photo effect etc.

\end{document}